\documentstyle[preprint,aps,graphicx]{revtex}
\tightenlines
\begin{document}
\title{$S=1$ kagom\'e Ising model with triquadratic interactions,
single-ion anisotropy and magnetic field: exact phase diagrams}
\author{J. H. Barry and K. A. Muttalib}
\address{Department of Physics, University of Florida,
P.O. Box 118440, Gainesville, FL 32611-8440}
\maketitle

\begin{abstract}
We consider a $S=1$ kagom\'e Ising model with triquadratic
interactions around each triangular face of the kagom\'e lattice,
single-ion anisotropy and an applied magnetic field. A mapping
establishes an equivalence between the magnetic canonical
partition function of the model and the grand canonical partition
function of a kagom\'e lattice-gas model with localized
three-particle interactions. Since exact phase diagrams are known
for condensation in the one-parameter lattice-gas model, the
mapping directly provides the corresponding exact phase diagrams
of the three-parameter $S=1$ Ising model. As anisotropy competes
with interactions, results include the appearance of confluent
singularities effecting changes in the topology of the phase
diagrams, phase boundary curves (magnetic field vs temperature)
with purely positive or negative slopes as well as intermediate
cases showing nonmonotonicity, and coexistence curves
(magnetization vs temperature) with varying shapes and
orientations, in some instances entrapping a homogeneous phase.\\


\end{abstract}

\newpage

{\bf I. Introduction}

The $S=1$ Ising model continues to be actively investigated with
applications in phase transitions, critical and multicritical
phenomena. The special behaviors of the model are largely
attributable to the presence of single-ion anisotropy and
biquadratic interaction terms in the Hamiltonian. A familiar
version is the Blume-Emery-Griffiths (BEG) model\cite{BEG} having
bilinear and biquadratic nearest-neighbor pair interactions and
single-ion-type uniaxial anisotropy. The three parameter BEG model
reduces to an earlier Blume-Capel model\cite{BC} if the
biquadratic interactions are neglected. The BEG model has
attracted considerable attention since it was originally proposed
to describe phase separation and superfluid ordering in
He$^3$-He$^4$ mixtures exhibiting tricritical behavior when the
anisotropy is sufficiently strong and competes with the
interactions. Subsequently, the model has been employed to explore
a variety of other phase transition problems\cite{Keskin}. Diverse
theoretical approaches have been adopted\cite{Keskin}, but
invariably due to severe mathematical difficulties, approximation
schemes are used in the calculations.

The free energy (characteristic function) of a lattice-statistical
model can be found in principle by evaluating the partition
function (``sum over states''), a formidable mathematical task in
a highly cooperative system with macroscopically many degrees of
freedom. The phase diagrams of the system indicate the locations
and other characteristics of the mathematical singularities in the
thermodynamic (large lattice) limit of the free energy per site,
$f$. Approaching the singularities, one examines whether the {\it
ordering parameter} (appropriate first-order derivatives of $f$)
vanishes continuously or jumps discontinuously as a function of
the temperature. The phase transition is then called continuous or
discontinuous, respectively, and is so signified upon the phase
diagrams.

In general, the phase diagrams of the $S=1$ BEG model are not
known exactly. However, neglecting its bilinear interactions,
Griffiths\cite{Griffiths} has pointed out that the partition
function of the remanent two-parameter $S=1$ Ising model is
equivalent to the partition function of a standard $S=1/2$ Ising
model in a field, whose exact phase diagrams are well-known in the
two-dimensional (d=2) ferromagnetic case. Wu\cite{Wu} extended the
investigations by considering a three-parameter $S=1$ Ising model
having a magnetic field, single-ion anisotropy and biquadratic
interactions, and proceeded to obtain exact phase diagrams of the
model for ferromagnetic interactions on planar lattices. In the
present paper, a $S=1$ kagom\'e Ising model is taken to have a
magnetic field, single-ion anisotropy and triquadratic
interactions around each triangular face of the kagom\'e lattice.
A mapping establishes an equivalence between the magnetic
canonical partition function of the model and the grand canonical
partition function of a kagom\'e lattice-gas model with localized
triplet interactions. Since exact phase diagrams are known for
condensation in the one-parameter lattice gas model, the mapping
directly provides the corresponding exact phase diagrams of the
three-parameter $S=1$ Ising model. As anisotropy competes with
interactions, special features of the findings include the
appearance of confluent singularities effecting changes in the
topology of the phase diagrams, phase boundary curves (magnetic
field vs temperature) with purely positive or negative slopes as
well as intermediate cases showing nonmonotonicity, and
coexistence curves (magnetization vs temperature) with varying
shapes and orientations, in some instances entrapping a
homogeneous phase. To our knowledge, the exact phase diagrams are
the first found for any $S=1$ planar Ising model with multi-spin
interactions.

As added incentive for these studies, it may be instructive to
briefly (albeit heuristically) discuss {\it multi-particle cyclic
exchange} processes. Letting ${\bf S}_l$ be a spin operator
localized on a lattice site $l$, an isotropic higher-order
Heisenberg exchange operator
\begin{equation}
({\bf S}_i\cdot {\bf S}_j)({\bf S}_j\cdot {\bf S}_k)({\bf
S}_k\cdot {\bf S}_i)\eqnum{1.1}
\end{equation}
can be associated with cyclic exchange around the vertex sites
$i$, $j$, $k$ of a triangle. Although not having the precise
electron spin operator forms (1.1), analogous multi-particle
exchange models may be relevant for particles with hard core
interactions where direct exchange is hindered at high densities
with the inference that the simplest exchange involves three
particles in a ring following one another in a circular motion. An
analogy can be experienced in an overcrowded bus or elevator where
it is easier for a few neighbors to revolve in unison than it is
for two neighbors to exchange their positions. Three-particle
cyclic exchanges are known to be important, e.g., in quantum
solids and liquids like helium\cite{Sullivan}. The highly
anisotropic or Ising version of (1.1) becomes
\begin{equation}
(S^z_i S^z_j)(S^z_j S^z_k)(S^z_k S^z_i)=
(S^z_i)^2(S^z_j)^2(S^z_k)^2, \eqnum{1.2}
\end{equation}
which is the operator form of the $S=1$ triquadratic Ising
interactions currently considered. The longitudinal Ising form
(1.2) is essentially the purely potential energy part of the
Heisenberg form (1.1). Although Ising model descriptions may not
be fully realistic, they can contain important vestiges of the
underlying complex problem, and offer valuable insights amid these
complexities. Exact solutions in the simplified models can
illustrate the role of multi-spin interactions in general and can
in particular highlight any possible novel effects of the
multi-spin interactions that may not be accessible or clearly
delineated using, e.g., perturbation theory, closed-form
approximations, or finite-size numerical simulations.

The paper is organized as follows. Section II presents the $S=1$
kagom\'e Ising model whose partition function is suitably
transformed using idempotent lattice-gas variables. In Section
III, the central mapping concepts in the theory are established.
Applying the mapping relations, phase boundary curves of the
present $S=1$ Ising model are determined in Section IV as are the
companion coexistence curves in Section V. Section VI contains
concluding comments on the theory and results.

{\bf II. Model Hamiltonian and partition function}

The kagom\'e lattice (Japanese woven bamboo pattern) is a
two-dimensional periodic array of equilateral triangles and
regular hexagons (Fig. 1) thus also called the 3-6 lattice. The
total numbers of triangles and hexagons are in a 2:1 ratio, with
the corner-sharing triangles having dual orientations, say, up- or
down-ward pointing. The lattice is regular (all sites equivalent,
all bonds equivalent) and may be termed ``close packed'' since it
contains elementary polygons having an odd number of sides, viz.,
triangles. Note that the kagom\'e lattice has the same
coordination number 4 as the square lattice, the latter being
``loose packed''.

Consider the following $S=1$ Ising (dimensionless) Hamiltonian
defined on the planar kagom\'e lattice of $\mathcal{N}$ sites
\begin{equation}
-\beta {\mathcal{H}} = h\sum_i S_i+D\sum_i
S^2_i+Q_3\sum_{<i,j,k>}S^2_i S^2_j S^2_k \eqnum{2.1}
\end{equation}
where $S_l=0,\pm 1$, $l=1,\cdots,\mathcal{N}$ are site-localized
spin-$1$ Ising variables, the summation $\sum_i$ is taken over all
lattice sites, $\sum_{<i,j,k>}$ is over all triplets of sites
belonging to elementary triangles, and $\beta=1/k_BT$ with $k_B$
being the Boltzmann constant and $T$ the absolute temperature.
Also, $h=\beta\tilde{\mu} h_z$ is a uniform (dimensionless)
magnetic field with $\tilde{\mu}$ being the electronic intrinsic
magnetic moment, $D=\beta\mathcal{D}$ a single-ion-type uniaxial
(dimensionless) anisotropy parameter, and $Q_3=\beta J_3>0$ a
three-spin (dimensionless) interaction parameter.  In (2.1), the
``symmetry breaking'' impressed field $h_z$ is longitudinal (along
z-axis) with the transverse $x-y$ plane taken as the plane of the
two-dimensional lattice. The magnetic field $h_z$ and the
anisotropy field $\mathcal{D}$ couple to the dipole and quadrupole
moments of the spin assembly, respectively. In the paper, the
lattice-statistical model (2.1) is termed a {\it S=1 triquadratic
Ising model}. The primary purpose of the theoretical
investigations is to deduce exact phase diagrams of the model by
employing transformation and mapping techniques upon its partition
function.

The magnetic canonical partition function ${\mathcal{Z}}(h,D,Q_3)$
of the spin system (2.1) is defined as
\begin{equation}
{\mathcal{Z}}(h,D,Q_3)=\sum_{\{S_i\}}e^{-\beta\mathcal{H}}
\eqnum{2.2}
\end{equation}
where the summation is taken over all possible values of the set
$\{S_i\}$ of spin variables. Explicitly entering (2.1), the
partition function (2.2) may be written as
\begin{eqnarray}
{\mathcal{Z}}(h,D,Q_3)&=&\sum_{\{S_i\}}e^{h\sum_iS_i+D\sum_iS^2_i
+Q_3\sum_{<i,j,k>}S^2_iS^2_jS^2_k}\cr &=&\sum_{\{n_i\}}(2\cosh
h)^{\sum_in_i} \times e^{D\sum_in_i+Q_3\sum_{<i,j,k>}n_in_jn_k}\cr
&=& \sum_{\{n_i\}}e^{[\ln(2\cosh
h)+D]\sum_in_i+Q_3\sum_{<i,j,k>}n_in_jn_k} \eqnum{2.3}
\end{eqnarray}
having made repeated use of the partial trace identity \cite{Wu}
\begin{equation}
\sum_{S_i=0,\pm 1}e^{hS_i}f(S^2_i)=\sum_{n_i = 0, 1}(2\cosh
h)^{n_i}f(n_i), \eqnum{2.4}
\end{equation}
which is readily established by identifying the terms
corresponding to $S_i=\pm 1(0)$ on the LHS with those
corresponding to $n_i=1(0)$ on the RHS. The enlistment of
idempotent variables $n_i$, $i=1,\cdots, {\mathcal{N}}$ in (2.3)
will be useful to directly establish a mapping between the
partition functions of the $S=1$ triquadratic  Ising model and a
triplet-interaction kagom\'e lattice gas model whose phase
diagrams for condensation are known exactly.

{\bf III. Lattice gas representation of $S=1$ triquadratic Ising
model}

Consider a lattice gas of $N$ atoms upon the kagom\'e lattice of
${\mathcal{N}}$ sites with the (dimensionless) Hamiltonian
\begin{equation}
-\beta{\mathcal{H}}_{lg}=K_3\sum_{<i,j,k>}n_in_jn_k, \eqnum{3.1}
\end{equation}
where $K_3=\beta\epsilon_3$ with $\epsilon_3>0$ being the strength
parameter of the short-range attractive triplet interaction, the
sum is taken over all elementary triangles, and the idempotent
site-occupation numbers are defined as
\begin{equation}
n_l=\cases{1,&site $l$ occupied \cr 0, & site $l$ empty.\cr}
\eqnum{3.2}
\end{equation}
In (3.1), an infinitely-strong (hard core) repulsive pair
potential has also been tacitly assumed for atoms on the {\it
same} site, thereby preventing multiple occupancy of any site as
reflected in the occupation numbers.

In the usual context of the grand canonical ensemble, we introduce
\begin{equation}
{\mathrm{H}}\equiv {\mathcal{H}}_{lg} -\mu N \eqnum{3.3}
\end{equation}
where $\mu$ is the chemical potential with $N$ being the conjugate
total number of particles
\begin{equation}
N=\sum_in_i\;\; . \eqnum{3.4}
\end{equation}
Using (3.1), (3.3) and (3.4), the grand canonical partition
function $\Xi(\mu,{\mathcal{N}},T)$ is given by
\begin{equation}
\Xi(\mu,{\mathcal{N}},T)=\sum_{\{n_i\}}e^{-\beta{\mathrm{H}}}
=\sum_{\{n_i\}}e^{\beta\mu\sum_in_i+K_3\sum_{<i,j,k>}n_in_jn_k}.
\eqnum{3.5}
\end{equation}
Comparing (2.3) and (3.5), we conclude that
\begin{equation}
{\mathcal{Z}}(h,D,Q_3)=\Xi(\mu,{\mathcal{N}},T) \eqnum{3.6a}
\end{equation}
provided that
\begin{equation}
\ln (2\cosh h)+D=\beta\mu, \;\;\; Q_3=K_3. \eqnum{3.6b}
\end{equation}
The phase diagrams of the triplet-interaction kagom\'e lattice gas
(3.1) are known exactly \cite{BS},more definitely, its
liquid-vapor phase boundary (chemical potential or pressure versus
temperature),coexistence curve (density versus temperature) and
various critical properties. Hence, as shown shortly, the mapping
(3.6) affords a convenient method for directly determining the
corresponding magnetic phase diagrams of the $S=1$ triquadratic
Ising model (2.1).

{\bf IV. Phase boundaries of the $S=1$ triquadratic Ising model}

Considering the kagom\'e lattice gas (3.1) with purely
three-particle interactions, its liquid-vapor {\it phase boundary
curve} is given by \cite{BS}
\begin{equation}
\mu/\epsilon_3=-K^{-1}_3 \ln [(e^{K_3}-1)^{2/3}-1], \;\;\; 0\leq
K_{3c}/K_3\leq 1, \eqnum{4.1}
\end{equation}
with $\mu/\epsilon_3$ being a reduced chemical potential and
$K_{3c}/K_3 (=T/T_c)$ a reduced temperature where $K_{3c} (\equiv
\epsilon_3/k_BT_c) = \ln [(2+\sqrt{3})^3+1] =
3.96992\cdot\cdot\cdot$. The curvilinear phase boundary begins at
zero temperature with $\mu/\epsilon_3 = -2/3$ and ends
(analytically) at a critical point whose coordinates are
$K_{3c}/K_3 = 1$, $\mu/\epsilon_3\equiv \mu_c/\epsilon_3 =
-0.64469\cdot\cdot\cdot$. At zero temperature, the phase boundary
curve $\mu/\epsilon_3$ vs $T/T_c$ has a zero slope in accordance
with the Clausius-Clapeyron equation and third law of
thermodynamics. Otherwise, its slope is {\it positive}, which is
more discernible at temperatures closely below the critical
temperature.

To similarly investigate the $S=1$ triquadratic Ising model (2.1),
one directly applies the correspondence relation (3.6b) to
expression (4.1) thereby yielding
\begin{equation}
\alpha=-\frac{x}{Q_{3c}}\ln\left\{\left[(e^{Q_{3c}/x}-1)^{2/3}-1\right]
2\cosh\left(\frac{Q_{3c}\xi}{x}\right)\right\},\;\;\; 0\leq x\leq
1, \eqnum{4.2a}
\end{equation}
where $$ \xi\equiv h/Q_3 (=\tilde{\mu} h_z/J_3);\;\;\; \alpha
\equiv D/Q_3 (={\mathcal{D}}/J_3);\;\;\; x\equiv Q_{3c}/Q_3
(=T/T_c);$$
\begin{equation} Q_{3c}\equiv J_3/k_BT_c=\ln
[(2+\sqrt{3})^3+1]=3.96992\cdot\cdot\cdot . \eqnum{4.2b}
\end{equation}
Regarding notations, the initial parameters $h, D, Q_3$ in (2.1)
are dimensionless through use of a thermal energy $k_BT$ scale.
Here, for graphical preferences, the associated parameters $\xi,
\alpha, x$ in (4.2) are reduced using an interaction energy $J_3$
scale.

Expression (4.2) is plotted in Figure 2 as a {\it phase boundary
surface} $x=x(\xi,\alpha)$, beginning at zero temperature $(x=0)$
as a wedge-shaped locus and terminating along a {\it critical
$(x=1)$ curve}. The phase boundary surface has a number of unusual
characteristics, leading to several novel features in the phase
boundaries and coexistence curves. In the following, we will
discuss the projections of the surface onto the $\xi-\alpha$ and
$\xi-x$ planes which will illustrate some of these special
features.

In Figure 3, isotherms of (4.2) in the range $0\leq x\leq 1$ are
plotted in $\xi-\alpha$ space (``field space''). The resulting
transition region may be viewed geometrically as the projection of
the phase boundary surface (Figure 2) in $\xi-\alpha-x$ space upon
the $\xi-\alpha$ plane. In particular, (4.2) shows, at $\xi=0$,
that the zero-temperature $(x=0)$ isotherm in Figure 3 has its
angular apex at $\alpha^{max}_0=-2/3$ and that the critical
$(x=1)$ isotherm exhibits a symmetric rounded maximum at
$\alpha^{max}_c=-Q^{-1}_{3c}\ln
\{2[(e^{Q_{3c}}-1)^{2/3}-1]\}=-0.81929\cdots.$ Observing Figure 3,
it is evident that the model only admits phase transitions for
anisotropy sufficiently competing with interactions
(${\mathcal{D}}\leq -(2/3)J_3$). The existence of surface
undulations in Figure 2 leads projectively to the existence of
{\it crossing points} among isotherms in Figure 3. For instance,
the zero temperature $(x=0)$ and critical temperature $(x=1)$
isotherms intersect one another at
$(\xi_{cross},\alpha_{cross})=(\pm 0.30171\cdots, -0.96837\cdots)$
as shown in Appendix A. Note that not all isotherms cross at a
single point, as shown in Figure 4 by magnifying the boxed region
of Figure 3. It is also clear from Figure 3 that for cases
$\alpha^{max}_c<\alpha<\alpha^{max}_0$, only non-critical
transitions exist, while for cases $\alpha
>\alpha^{max}_0$, the model is devoid of any phase transition.

In practice, a magnetic field parameter $\xi$ is experimentally
adjustable, contrasting the intrinsic crystal-field anisotropy
parameter $\alpha$. This makes the projection of Figure 2 onto the
$\xi-x$ plane more relevant from an experimental point of view. As
an example, for a given specimen, say $\alpha=-1.5$, and for
$\xi>0$, the relation (4.2) shows that the phase transitions only
occur within the range $0.8333\cdots\leq \xi \leq 0.8550\cdots$,
where the lower and upper end points of the interval correspond to
the zero and critical temperatures, respectively. Since the
interval does not contain vanishing values of $\xi$, the
associated phase transitions for the specimen are {\it induced} by
the applied magnetic field $h_z$. For this example, the relation
(4.2) is plotted in Figure 5, illustrating the phase boundary
(solid curve) to be curvilinear with a {\it positive} slope at
non-zero temperatures. The phase boundary separates two
homogeneous phases, viz., the longitudinal upper phase associated
with the Ising spins preferably aligning parallel to the applied
field $h_z$ (along the positive $z$-axis) and the transverse lower
phase associated with the spins preferably lying in the $x-y$
plane of the lattice. The phase boundary is the locus of
discontinuous transitions from one phase to the other, and the
terminating point (solid circle) is a critical point associated
with a continuous transition. In Figure 5, the phase boundary is
analytic at the critical point. In fact, the present phase
boundary curves (magnetic field vs temperature) are analytic at
their critical points for all $\alpha < \alpha^{max}_c$. As
explained shortly, the phase boundary curve for $\alpha =
\alpha^{max}_c$ possesses a {\it confluent singularity} at
criticality $(x=1)$.

The {\it monotonically increasing} behavior of the phase boundary
curve in Figure 5, however, is not realized for all values of the
single-ion anisotropy parameter $\alpha$. Figure 6 shows the
monotonically increasing phase boundary curve for $\alpha \lesssim
-1$ gradually crossing over to a {\it monotonically decreasing}
phase boundary curve for $\alpha >\tilde{\alpha}\approx -.93724$.
In particular, selecting the value $\alpha_{cross}=-0.96837\cdots$
from within this crossover range, the resulting phase boundary
curve in Figure 7 exhibits {\it non-monotonic} behavior, viz., a
shallow rounded {\it minimum} having coordinates
($x_{min},\xi_{min})=(0.80577\cdots,0.30020\cdots)$. This
non-monotonic behavior of the phase boundary curves is associated
with the region of crossing points in Figure 3 (see Appendix A).

Upon further increasing the values of $\alpha$, the corresponding
$\xi$-coordinates of the critical points (solid circles) descend
in value, as shown in Figure 8, until the parameter value
$\alpha=\alpha^{max}_c=-0.81929\cdots$, when the phase boundary
curve vanishes as $A_{\xi}\epsilon^{1/2}$ (algebraic branch point
singularity). Here, $\epsilon\equiv(T_c-T)/T_c$ is the fractional
deviation of temperature from its critical value and the
calculated amplitude $A_{\xi}=0.20963\cdots.$ This singular
behavior arises as the critical point ($\xi>0$) shown in Figure 8
meets its reflective image critical point ($\xi<0$) at a {\it
confluence point} ($x,\xi$)=($1,0$), thereby changing the
topological nature of the phase diagram. Continuing to increase
$\alpha$ beyond its $\alpha^{max}_c$ value has the effect of
further shortening the length of the phase boundary curve
(comprising solely non-critical transitions) whose confluent
singularity moves toward the origin, carrying the same exponent
$1/2$ but diminishing amplitude. Finally, as expected, the phase
boundary curve no longer exists for $\alpha > -2/3$.

{\bf V. Coexistence curves of the $S=1$ triquadratic Ising model}

In the previous section, mapping relations were directly employed
to facilitate the finding of exact phase boundary curves of the
$S=1$ triquadratic Ising model. In the present section, the
companion coexistence curves of the model are determined via
composite differentiation of the same mapping relations (3.6).
More particularly, the (dimensionless) {\it magnetization} $m$ is
represented as
\begin{eqnarray}
m\equiv <S_i> &=& {\mathcal{N}}^{-1}\frac{\partial \ln
{\mathcal{Z}}(h,D,Q_3)}{\partial h}\cr &=&
{\mathcal{N}}^{-1}\frac{\partial \ln
\Xi(\mu,{\mathcal{N}},T)}{\partial (\beta\mu)}\cdot \frac{\partial
(\beta\mu)}{\partial h}\cr &=& <n_i> \cdot\frac{\partial [\ln (2
\cosh h)+D]}{\partial h}\cr &=& \rho \cdot \tanh h \eqnum{5.1}
\end{eqnarray}
where $\rho\equiv <n_i>$ is the {\it average particle number
density} of the triplet interaction kagom\'e lattice gas (3.1). In
the context of coexistence phase diagrams, the correspondence
relationship (5.1) specializes to
\begin{equation}
m^{coex}_{l,t}= \rho^{coex}_{l,v}(\tanh h)^{pbs} \eqnum{5.2}
\end{equation}
where $m^{coex}_{l,t}$ is the {\it longitudinal-transverse
coexistence surface} in $m-\alpha-x$ space of the $S=1$
triquadratic Ising model, $\rho^{coex}_{l,v}$ is the {\it
liquid-vapor coexistence curve} in $\rho-x$ space of the triplet
interaction kagom\'e lattice gas and $(\tanh h)^{pbs}$ is
evaluated upon the earlier determined phase boundary surface (pbs)
(4.2), the latter yielding
\begin{eqnarray}
(\tanh h)^{pbs} &=& \left(\sqrt{1-\cosh^{-2}h}\;\right)^{pbs}\cr
&=& \left\{1-4e^{2\alpha\frac{Q_{3c}}{x}}\left[
(e^{\frac{Q_{3c}}{x}}-1)^{2/3}-1\right]^2\right\}^{1/2},\;\;\;
0\leq x \leq 1. \eqnum{5.3}
\end{eqnarray}
In the product representation (5.2), the $\alpha$-dependence of
$m^{coex}_{l,t}$ resides solely in the factor $(\tanh h)^{pbs}$ as
explicitly given in (5.3). For a {\it fixed} value of the
anisotropy parameter $\alpha$, the expression (5.3) evaluates
$(\tanh h)^{pbs}$ along the associated phase boundary curve
(magnetic field vs temperature).

The exact solution for the liquid-vapor coexistence curve
($\rho^{coex}_{l,v}$ vs temperature) of the triplet interaction
kagom\'e lattice gas (3.1) is given by\cite{BS}
\begin{equation}
\rho^{coex}_{l,v}=1-\frac{1}{4}\left[1-(e^{K_3}-1)^{-2/3}\right]
\left[1+<\mu_0\mu_1>_{L^*=0}\mp 2<\mu>_S\right], \;\;\; 0\leq
K_{3c}/K_3\leq 1, \eqnum{5.4}
\end{equation}
where $<\mu_0\mu_1>_{L^*=0}$, $<\mu>_S$ are the nearest-neighbor
pair correlation and spontaneous magnetization, respectively, in a
standard $S=\frac{1}{2}$ honeycomb Ising model ferromagnet with
nearest-neighbor pair (dimensionless) interaction parameter $K^* >
0$ and (dimensionless) magnetic field $L^*$. It is well known
\cite{Griffiths2} that a necessary and sufficient condition for
the existence of a phase transition in the $S=\frac{1}{2}$ {\it
ferromagnetic} ($K^*>0$) honeycomb Ising model is the joint
condition $L^*=0$ and $K^*\geq K^*_c$, where the critical value
$K^*_c=\frac{1}{2}\ln (2+\sqrt{3})=0.65847\cdots$. In (5.4), the
exact solutions for $<\mu_0\mu_1>_{L^*=0}$, and $<\mu>_S$ are
known to be \cite{Houtappel}
\begin{equation}
<\mu_0\mu_1>_{L^*=0}=\frac{2}{3}\left[\coth 2K^*+\gamma
K_1(\kappa)\right], \eqnum{5.5a}
\end{equation}
\begin{equation}
\langle \mu \rangle_S = \cases{(1-\kappa^2)^{1/8}, &$0\le
K^*_c/K^* \le 1$, \cr 0,&$1< K^*_c/K^*$}\eqnum{5.5b}
\end{equation}
with $K_1(\kappa)$ being the complete elliptic integral of the
first kind
\begin{equation}
K_1(\kappa)=\int_0^{\pi/2}(1-\kappa^2 \sin^2\theta)^{-1/2}d\theta,
\eqnum{5.6a}
\end{equation}
and where
\begin{equation}
\kappa^2=16z^3(1+z^3)(1-z)^{-3}(1-z^2)^{-3}, \eqnum{5.6b}
\end{equation}
\begin{equation}
\gamma=(1-z^4)(z^2-4z+1)/\pi|1-z^2|(1-z)^4, \eqnum{5.6c}
\end{equation}
\begin{equation}
z=e^{-2K^*}. \eqnum{5.6d}
\end{equation}
The expression (5.4) for $\rho^{coex}_{l,v}$ can be written
completely in the natural $K_3$-notation of the lattice gas by
substituting the interaction parameter relation \cite{BS}
\begin{equation}
K^*=\frac{1}{6}\ln (e^{K_3}-1)\;\;\;\hbox{at} \;\;\; L^*=0
\eqnum{5.7}
\end{equation}
into (5.5) and (5.6). Also, the earlier stated critical value
$K_{3c}=\ln [(2+\sqrt{3})^3+1]=3.96992\cdots$ is determined by
substituting the known critical value $K^*_c=\frac{1}{2}\ln
(2+\sqrt{3})$ into (5.7).

The goal of the present section has been attained. Namely, for a
given value of the single-ion anisotropy parameter $\alpha$, the
{\it longitudinal-transverse coexistence curve} ($m^{coex}_{l,t}$
vs temperature) of the $S=1$ triquadratic Ising model is exactly
calculable in $m-x$ space by substituting (5.3-7) into (5.2) and
replacing $K_3$ whenever appearing by $Q_3$ in compliance with the
mapping relation (3.6b). The resulting coexistence curves display
novel behaviors as illustrated in Figures 9, 10 and 11.

In Figure 9, the longitudinal-transverse coexistence curve is
plotted for $\alpha=-1.5$, exhibiting an asymmetric rounded shape.
For increasing temperatures, the transverse (lower) branch
$m^{coex}_t$ rises more rapidly than the longitudinal (upper)
branch $m^{coex}_l$ falls, causing asymmetry in the rounded shape
of the coexistence curve. The resulting curvilinear diameter of
the coexistence region (solid line) increases monotonically, which
is more pronounced at temperatures closely below the critical
temperature (the diameter of the coexistence region is defined as
the arithmetic mean $\frac{1}{2}(m^{coex}_l+m^{coex}_t)$ of the
longitudinal and transverse branches of the coexistence curve or,
geometrically, the locus of the midpoints of the vertical
``tie-lines'' spanning the coexistence region). The coexistence
curve in Figure 9 is singular at its critical point (solid
circle). This singular behavior originates within the factor
$\rho^{coex}_{l,v}$ of the product representation (5.2).
Specifically, upon inspecting (5.4), the coexistence curve
superposes the algebraic branch point ($\epsilon^{1/8}$) and weak
energy-type ($\epsilon\ln \epsilon$) singularities carried by
$<\mu>_S$ and $<\mu_0\mu_1>_{L^*=0}$, respectively, resulting in
an infinite (vertical) slope at its critical point.

Increasing values of the anisotropy parameter $\alpha$ alter the
shape and orientation of the coexistence  curve. Including the
previous case $\alpha=-1.5$, Figures 10 and 11 exhibit coexistence
curves and their curvilinear diameters, respectively, for
increasing values of $\alpha$. To illustrate, $\alpha = -0.87$
reveals an asymmetric rounded coexistence curve (curve 2 in Figure
10) with a {\it monotonically decreasing} curvilinear diameter
(curve 2 in Figure 11), and a narrowing of the coexistence region
within the temperature interval $0.5 \lesssim x \leq 1$, where the
upper (longitudinal) branch of the coexistence curve falls more
rapidly. As argued earlier for Figure 9, the coexistence curve 2
in Figure 10 is singular at its critical point (solid circle) with
the leading singularity again being an algebraic branch point
($\epsilon^{1/8}$).

As $\alpha$ ascends in value, the $m$-coordinates of the critical
points (solid circles) descend in value until vanishing for the
value $\alpha=\alpha^{max}_c=-0.81929\cdots$ (curve 3 in Figures
10 and 11). Here the critical point ($m > 0$) meets its reflective
image critical point ($m < 0$) at a confluence point
$(x,m)=(1,0)$, thereby changing the topology of the phase diagram.
Note then that the transverse branches of the coexistence curves
entrap a homogeneous transverse phase within a tadpole shape
region whose positive portion ($m>0$) appears in Figure 10. Also,
the analytical behavior of the coexistence curve near its
confluence point $(x,m)=(1,0)$ deserves careful scrutiny since
{\it both} factors in the product representation (5.2) are
singular. Sufficiently close to the confluence point, the factor
$\rho^{coex}_{l,v}=\rho_c+o(\epsilon^{1/8})$ where the critical
density
$\rho_c=\frac{1}{3}(\frac{7}{2}-\sqrt{3})=0.58931\cdots$\cite{BS},
and using (4.2b), the confluence factor $(\tanh h)^{pbs}$ is
evaluated along the $\alpha=\alpha^{max}_c$ phase boundary curve
(pbc) as $(\tanh h)^{pbs}=(\tanh \frac{Q_{3c}\xi}{x})^{pbc}
=Q_{3c}A_{\xi}\epsilon^{1/2}+ o(\epsilon^{3/2})$, as discussed in
Section IV.  Hence, substituting the above expansions into the
product representation (5.2), one concludes that, closely
approaching its confluence point, the $\alpha=\alpha^{max}_c$
coexistence curve vanishes as $A_m\epsilon^{1/2}$ (algebraic
branch point singularity) where the amplitude $A_m=\rho_c
Q_{3c}A_{\xi}=0.49043\cdots$, substituting known values of the
constants.

Continuing to increase $\alpha$ beyond its $\alpha^{max}_c$ value
further shortens the length of both the coexistence curve (Figure
10) and the corresponding curvilinear diameter (Figure 11) whose
confluent singularities (open circles) move toward the origin,
carrying the same exponent $1/2$ and, suggestively from Figure 10,
increasing (decreasing) amplitude for the upper (lower) branch of
the coexistence curve. Concurrently, the region encasing the
homogeneous transverse phase shrinks in size as it shifts toward
the origin. Eventually, as anticipated, phase diagrams for the
model are nonexistent if $\alpha > -2/3$.

The following concepts further elucidate the nature of the
confluent phase transitions. Since the magnetization order
parameter (length of vertical tie-line spanning the coexistence
region in Figure 10) vanishes continuously at a confluence point,
the phase transition is a continuous type. One can also argue that
the confluence point (open circle) is non-critical since it is a
{\it non-vertical} inflection point of the $x=x_{confl}$
isothermal curve in the $m$ (magnetization) vs $\xi$ (magnetic
field) plane. In other words, the initial isothermal magnetic
susceptibility remains finite at the confluence point (open
circle), contrasting the case of a diverging susceptibility which
is a thermodynamic hallmark of a magnetic critical point (solid
circle). The ``hybrid'' singular point (shaded circle) is also
associated with a divergent susceptibility as discussed in
Appendix B.

In the phase diagrams of Sections IV and V, we have chosen to
accentuate the dipolar stimulus $\xi$ and the dipolar thermal
response $<S_i>$, both being readily measurable quantities.
Comparatively, the quadrupolar quantities $\alpha$ and $<S^2_i>$
are not as easily adjustable or accessible experimentally.
Theoretically, however, (4.2) directly yields the phase boundary
curves $\alpha$(anisotropy) vs $x$(temperature) for fixed values
of $\xi$(magnetic field). Simple geometrical inspection of Figure
3 reveals that these quadrupolar phase boundary curves are devoid
of confluent singularities and that each curve ends at a critical
point. Also, composite differentiation of the mapping relations
(3.9) provides the correspondence relationship $<S^2_i>=\rho$,
leading to a {\it solitary} coexistence curve
$\left<S^2_i\right>^{coex}_{l,t}$ vs $x$(temperature), depending
{\it only} upon the temperature in the transition region of Figure
3.

Lastly, one notes that the single-spin configurational
probabilities $p_+$, $p_0$, $p_-$, in obvious notations, may be
represented in terms of the thermal averages $<S_i>$, $<S^2_i>$
(ordering parameters) as $p_{\pm}=\frac{1}{2}(<S^2_i>\pm <S_i>)$,
$p_0=1-<S^2_i>$. Hence, within the current theoretical framework,
these configurational probabilities can be exactly evaluated along
the coexistence curves (or across the associated phase boundary
curves). Indeed, the resulting configurational probabilities
confirm earlier descriptions of the $S=1$ spin orientations as
preferentially longitudinal (transverse) in the upper (lower)
phases of the diagrams.

{\bf VI. Concluding Remarks}

Exact results in physics are valuable for a variety of reasons.
Endeavoring to retain only the most essential ingredients of a
physical problem, exact solutions of simple model systems often
provide definite guidance and insights on more realistic and
invariably more mathematically complex systems. Exact results in
tractable models of seemingly different physical systems may alert
researchers to significant common features of these systems and
actually emphasize concepts of universality. In addition to their
own aesthetic appeal, exact results can, of course, serve as
standards against which both approximation methods and approximate
results may be appraised. Also, the underlying mathematical
structures of exactly soluble models in statistical physics are
rich in content and have led to important developments in
mathematics.

In the present theoretical investigations, a $S=1$ kagom\'e Ising
model was taken to have localized triquadratic interactions
($J_3$), single-ion anisotropy ($\mathcal{D}$), and was placed in
a uniform magnetic field $h_z$. As anisotropy sufficiently
competes with interactions (${\mathcal{D}} \leq -2J_3/3$), the
model admits phase transitions and its phase diagrams were
determined using transformation and mapping methods upon its
partition function. The mapping techniques significantly
simplified the calculations. More particularly, since exact phase
diagrams are known for condensation in a one-parameter kagom\'e
lattice gas model with triplet interactions\cite{BS}, the mapping
directly afforded the corresponding exact magnetic phase diagrams
of the three-parameter $S=1$ kagom\'e Ising model. Otherwise, as
previously demonstrated \cite{BS} with the aforementioned lattice
gas model, the theory would incorporate a symmetric eight-vertex
model on the honeycomb lattice in a mediating role and the
computations then lengthen considerably.

Special features in the results included the appearance of
confluent singularities causing changes in the topology of the
phase diagrams, phase boundary curves (magnetic field vs
temperature) with purely positive or negative slopes as well as
intermediate cases showing nonmonotonicity, and coexistence curves
(magnetization vs temperature) with varying shapes and
orientations, in some instances entrapping a homogeneous phase.
The phase diagrams indicate both discontinuous and continuous
transitions, the latter being confluent type (open circles),
confluent-critical type (shaded circles), and critical type (solid
circles). More explicitly, as a function of the temperature, the
magnetization order parameter vanishes with exponent $1/2$ at a
confluent or confluent-critical singularity, and with exponent
$1/8$ at a critical singularity. Also, the magnetic susceptibility
diverges at the confluent-critical point with an exponent
$\gamma'=3/4$.

Towards these ends, the phase boundary surface in Figure 2
(temperature as a function of the magnetic and anisotropy fields)
was projected upon the field plane in Figure 3 showing the
isotherms $0\leq T \leq T_c$ in the plane to possess {\it crossing
points}. Such a region of crossing points for finite fields does
not occur in the model having {\it biquadratic} interactions
studied by Wu \cite{Wu}. Hence, some features found in the present
phase diagrams may be deemed {\it multi-spin} interaction effects.
As examples, both the purely positive and nonmonotonic slopes of
the phase boundary curves in Figures 5-7 are connected with the
presence of crossing points in Figure 3 and their originating {\it
triquadratic} interactions in the model Hamiltonian (2.1). To our
knowledge, the exact phase diagrams of the present $S=1$
triquadratic Ising model are the first obtained for any $S=1$
planar Ising model with multi-spin interactions.\\

{\bf  Acknowledgment}

The authors are grateful to the referee for valuable comments and
suggestions.

{\bf Appendix A: Crossing points of the $x=0$ isotherm with the
$x=1$ isotherm in the $\xi-\alpha$ plane}

Consider first the $\xi>0$ portion of the transition region in
Figure 3. As $x\rightarrow 0$ ($T\rightarrow 0$), the leading
exponential behaviors in (4.2a) provide the {\it zero temperature
isotherm} as
\begin{eqnarray}
\alpha \rightarrow \alpha_0&=&-\frac{x}{Q_{3c}}\ln
\left(e^{2Q_{3c}/3x}e^{\xi Q_{3c}/x}\right) \cr
&=&-\frac{x}{Q_{3c}}\left[\frac{2Q_{3c}}{3x}+\frac{\xi
Q_{3c}}{x}\right] \cr &=&-\frac{2}{3}-\xi. \eqnum{A.1}
\end{eqnarray}
At criticality, the temperature variable $x=1$ $(T=T_c)$ and
(4.2a) affords the {\it critical isotherm} as
\begin{eqnarray}
\alpha=\alpha_c&=&-\frac{1}{Q_{3c}}\ln
\left(\left[(e^{Q_{3c}}-1)^{2/3}-1\right] 2 \cosh (Q_{3c}\xi)
\right) \cr &=&-\frac{1}{Q_{3c}}\ln\left[4(3+2\sqrt{3})\cosh
(Q_{3c}\xi)\right], \eqnum{A.2}
\end{eqnarray}
having substituted $(e^{Q_{3c}}-1)^{2/3}-1=2(3+2\sqrt{3})$ via
(4.2b). To locate their \textit{crossing point}, one equates above
expressions $\alpha_0=\alpha_c$ yielding
$$
e^{(\xi_{cross}+2/3)Q_{3c}}=4(3+2\sqrt{3})\cosh
(Q_{3c}\xi_{cross})
$$
or
\begin{equation}
\xi_{cross}=(2Q_{3c})^{-1}\ln
\left[\frac{2(3+2\sqrt{3})}{e^{2Q_{3c}/3}-2(3+2\sqrt{3})}\right]
\eqnum{A.3}.
\end{equation}
Substituting (4.2b), expression (A.3) gives
\begin{equation}
\xi_{cross}=0.12594\cdots\ln\frac{12.92820\cdots}{1.17806\cdots}=0.30171\cdots,
\eqnum{A.4}
\end{equation}
which is substituted into (A.1) yielding
\begin{equation}
\alpha_{cross}=-2/3-\xi_{cross}=-0.96837\cdots. \eqnum{A.5}
\end{equation}
Similar calculations for the $\xi<0$ portion of the transition
region in Figure 3 (or by simply recognizing the even symmetry of
the $\xi$-dependence in (4.2a)) determine the coordinates of the
crossing points
\begin{equation}
\xi_{cross}=\pm 0.30171\cdots, \;\;\;
\alpha_{cross}=-0.96837\cdots. \eqnum{A.6}
\end{equation}
In the vicinity of the crossing points (A.6), countless other
crossing points occur among the isotherms as shown in Figure 4.\\

{\bf Appendix B: Divergent susceptibility ($\gamma'=3/4$) at the
confluent-critical point}

The correspondence relation (5.1) is written
\begin{equation}
m=\tilde{\mu}\cdot\rho \cdot \tanh h, \eqnum{B.1}
\end{equation}
re-entering the electronic intrinsic magnetic moment
$\tilde{\mu}$. Differentiating (B.1) with respect to the magnetic
field provides an additional correspondence relation between
thermodynamic response coefficients of the $S=1$ triquadratic
Ising model and the triplet-interaction kagom\'e lattice gas
model. Attention will ultimately be directed toward the ``hybrid''
singular points (shaded circles) in Figure 8 and Figure 10.

Specifically, the {\it initial isothermal magnetic susceptibility}
$\chi_T$ of the $S=1$ triquadratic Ising model is defined (in
appropriate units) by
\begin{eqnarray}
\chi_T\equiv \lim_{h_z\rightarrow 0}\frac{\partial m}{\partial
h_z} &=& \beta\tilde{\mu}\lim_{h\rightarrow 0}\frac{\partial
m}{\partial h} \cr &=& \beta\tilde{\mu}^2\lim_{h\rightarrow
0}\frac{\partial (\rho\cdot\tanh h)}{\partial h}\cr &=&
\beta\tilde{\mu}^2\lim_{h\rightarrow 0}\left[\frac{\partial
\rho}{\partial (\beta\mu)}\frac{\partial (\beta\mu)}{\partial
h}\tanh h+\rho\frac{\partial \tanh h}{\partial h}\right]\cr &=&
\tilde{\mu}^2\lim_{h\rightarrow 0}\left[\frac{\partial
\rho}{\partial \mu}\tanh^2 h+\beta\rho/ \cosh^2 h\right],
\eqnum{B.2}
\end{eqnarray}
having used (B.1), the mapping relation (3.6b) and the
(dimensionless) field parameter $h=\beta\tilde{\mu}h_z$. In (B.2),
all partial derivatives are performed at constant temperature.

The grand canonical partition function of the triplet-interaction
kagom\'e lattice gas model is equivalent (aside from known
pre-factors) to the magnetic canonical partition function of a
standard $S=1/2$ honeycomb Ising model with pair interactions and
field \cite{BS}. The thermal behaviors of the models exhibit each
to have a single critical point with corresponding critical
exponents being equal. The exponent equivalence will be used
shortly regarding the critical exponent $\gamma'=7/4$. The {\it
isothermal compressibility} $\kappa_T$ of the triplet-interaction
kagom\'e lattice gas can be found (in appropriate units) from
\begin{equation}
\kappa_T=\frac{1}{\rho^2}\left(\frac{\partial \rho}{\partial
\mu}\right)_T, \eqnum{B.3}
\end{equation}
which in turn is substituted into (B.2) yielding
\begin{equation}
\chi_T=\tilde{\mu}^2\lim_{h\rightarrow
0}\left[\rho^2\kappa_T\tanh^2 h+\beta\rho/ \cosh^2 h\right]
\eqnum{B.4},
\end{equation}
the sought relationship between response coefficients $\chi_T$ and
$\kappa_T$.

Consider, in particular, the singular point (shaded circle) in
Figure 10 having a confluence of critical $(x=1)$ points for
$\alpha=\alpha^{max}_c$. As $x\rightarrow 1^{-}$ $(T\rightarrow
T^{-}_c)$ along the $\alpha=\alpha^{max}_c$ coexistence curve, one
argues sufficiently near the singular point that
$\kappa_T\rightarrow\infty$ as $\epsilon^{-7/4}$ in (B.4), and
recalls that $\tanh^2 h\rightarrow 0$ as $\epsilon$ along the
associated phase boundary curve in Figure 8. Hence the
relationship (B.4) directly implies that $\chi_T\rightarrow\infty$
as $\epsilon^{-3/4}$. In effect, the strongly divergent
$(\gamma'=7/4)$ susceptibility found at criticality in a standard
$S=1/2$ planar Ising model is weakened $(\gamma'=3/4)$ at the
confluence of critical points in the $S=1$ triquadratic Ising
model. Additionally, the relationship (B.4) verifies that $\chi_T$
remains finite at the other confluence points (open circles) in
Figure 10.

\begin{figure}
\begin{center}
\leavevmode\includegraphics[width=0.5\textheight,angle=-90]{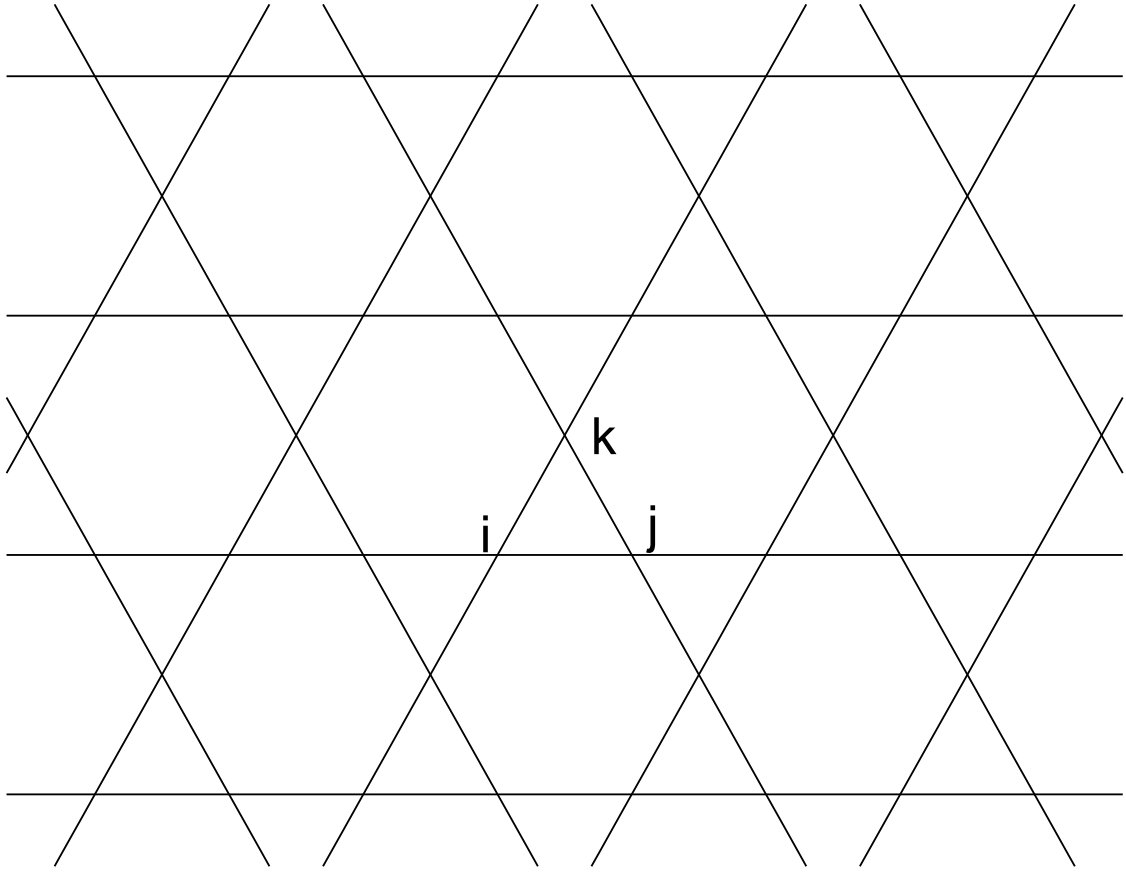}
\end{center}
\caption{The kagom\'e lattice is a two-dimensional periodic array
of equilateral triangles and regular hexagons. Sites $i,j,k$ are
vertices of an elementary triangle.}
\end{figure}

\begin{figure}
\begin{center}
\leavevmode\includegraphics[width=0.7\textheight]{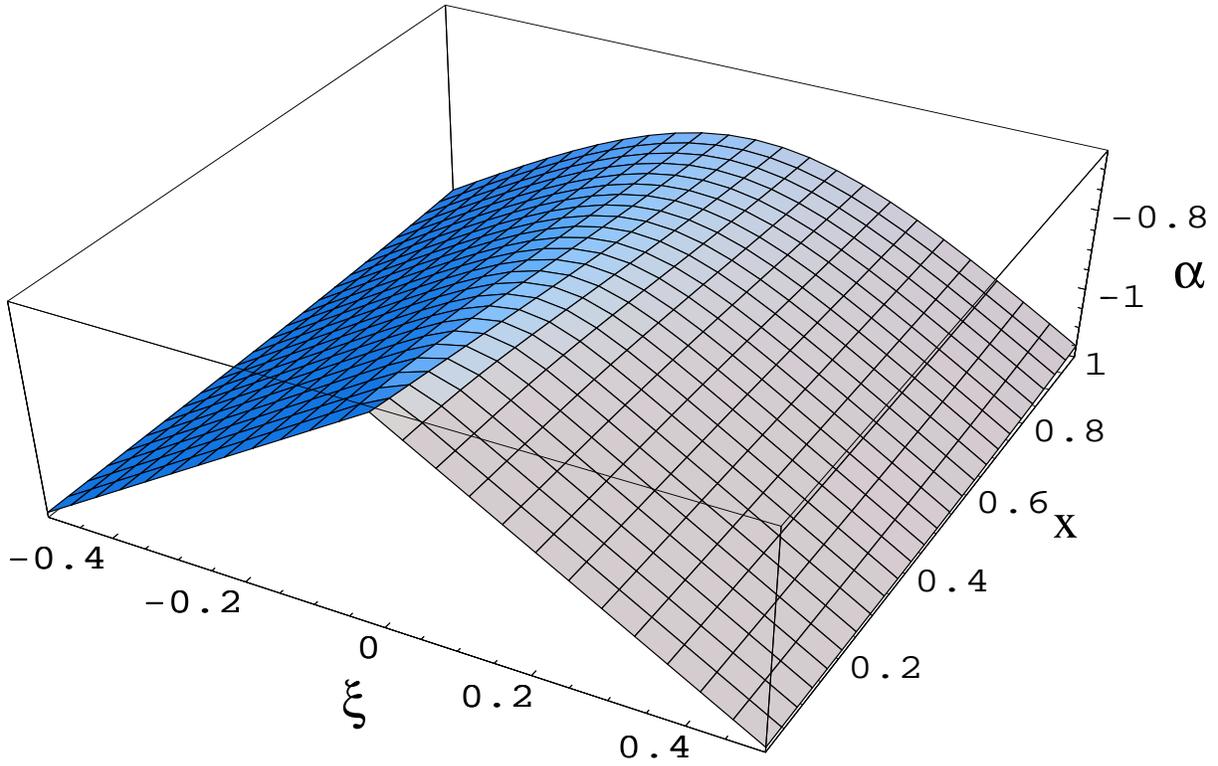}
\end{center}
\caption{Phase boundary surface $x=x(\xi,\alpha)$ where $x$ is the
reduced temperature, $\xi$ is the reduced magnetic field and
$\alpha$ is the reduced anisotropy parameter. Figures 3-8 are
projections of this surface onto the $\xi - \alpha$ and $\xi - x$
planes and will illustrate several unusual features of the
surface. The reduced parameters $\xi=\tilde\mu h_z/J_3$,
$\alpha={\mathcal{D}}/J_3$, $x=T/T_c$ where
$k_BT_c/J_3=Q^{-1}_{3c}=0.25189\cdots$.}
\end{figure}
\newpage

\begin{figure}
\begin{center}
\leavevmode
\includegraphics[width=0.5\textheight,angle=-90]{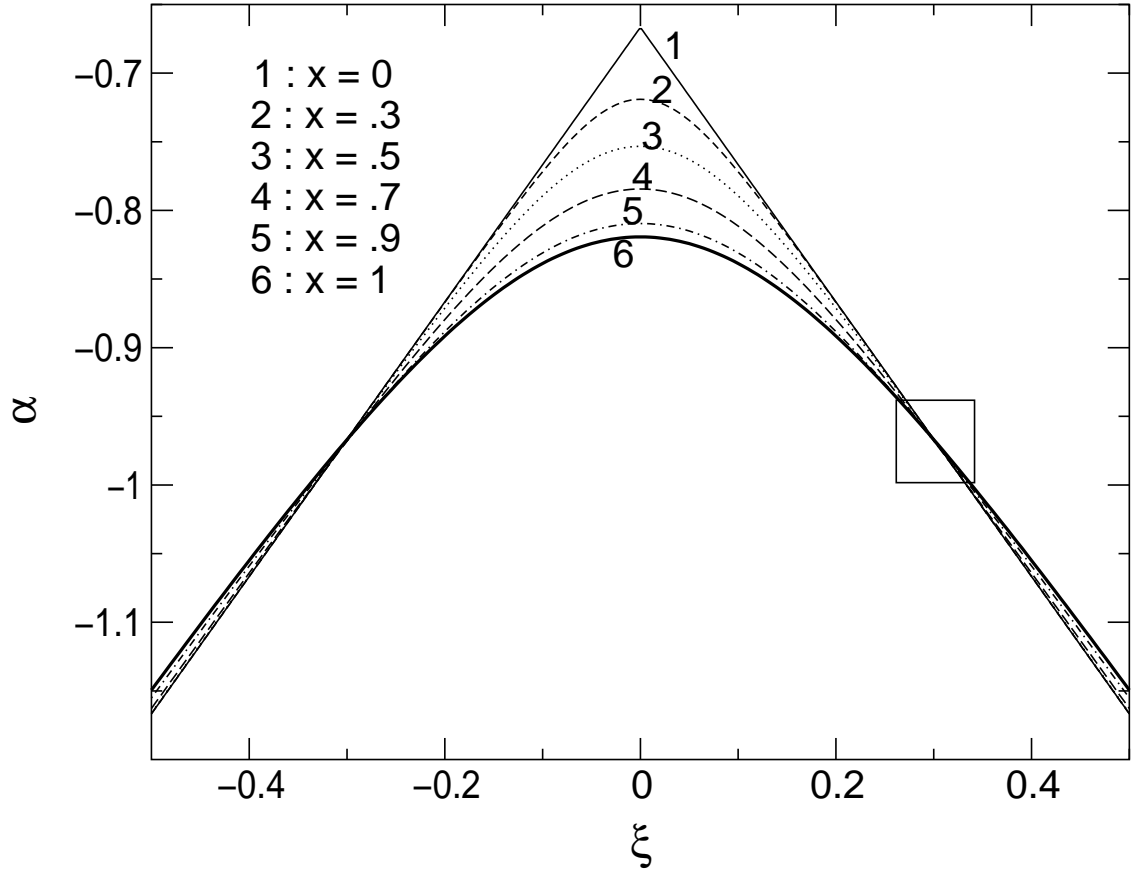}
\end{center}
\caption{Isotherms in the $\xi$(magnetic field) $-$
$\alpha$(anisotropy) plane for several values of $x$(temperature),
including the critical isotherm $x = 1$. The existence of surface
undulations in Figure 2 leads to the crossing of isotherms (boxed
region). The reduced parameters $\xi=\tilde\mu h_z/J_3$,
$\alpha={\mathcal{D}}/J_3$, $x=T/T_c$ where
$k_BT_c/J_3=Q^{-1}_{3c}=0.25189\cdots$.}
\end{figure}

\begin{figure}
\begin{center}
\leavevmode
\includegraphics[width=0.5\textheight,angle=-90]{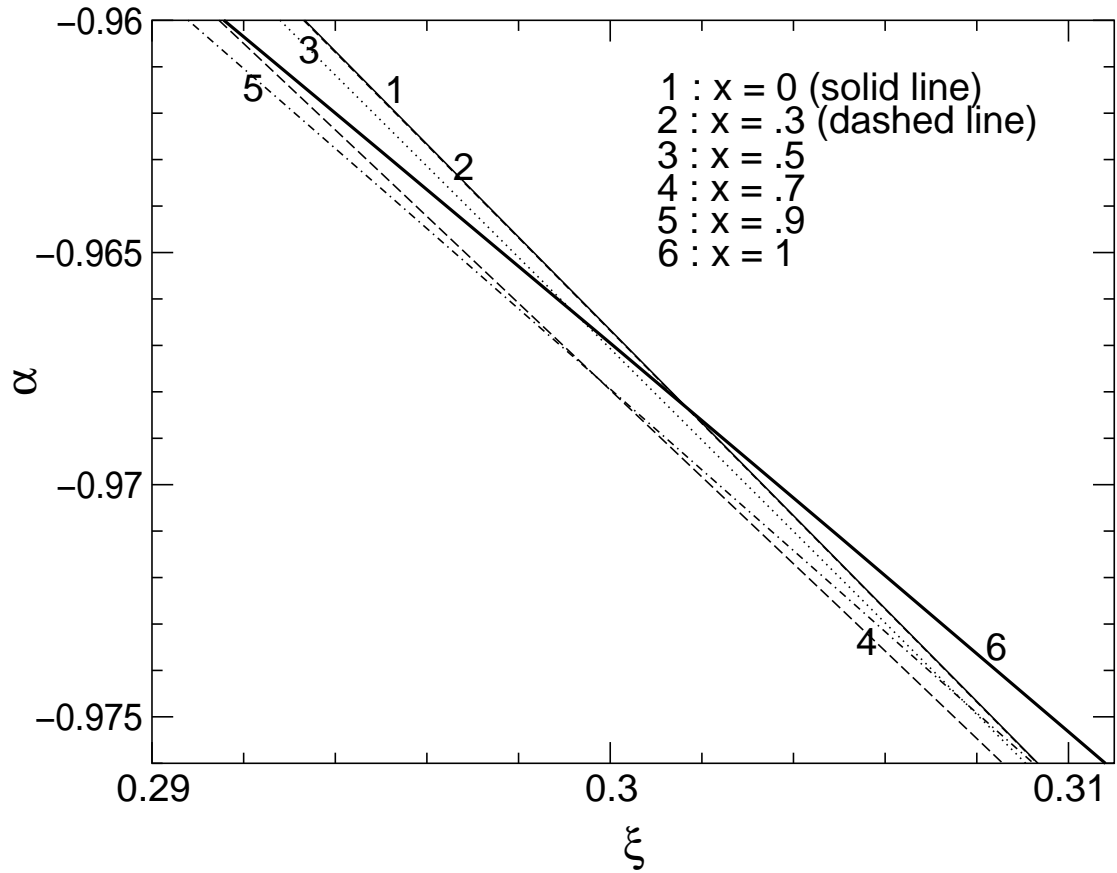}
\end{center}
\caption{Boxed region in Figure 3  magnified to show details. The
reduced parameters $\xi=\tilde\mu h_z/J_3$,
$\alpha={\mathcal{D}}/J_3$, $x=T/T_c$ where
$k_BT_c/J_3=Q^{-1}_{3c}=0.25189\cdots$.}
\end{figure}

\begin{figure}
\begin{center}
\leavevmode
\includegraphics[width=0.5\textheight,angle=-90]{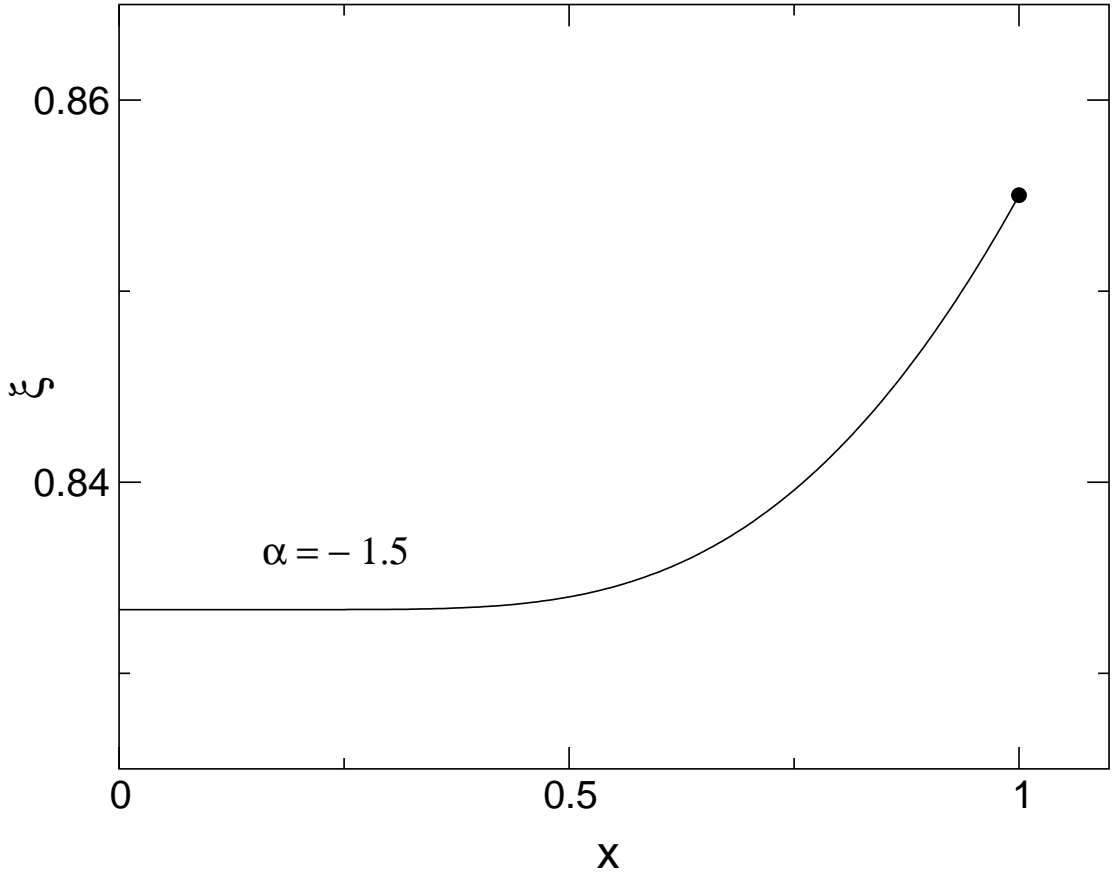}
\end{center}
\caption{Projection of Figure 2 onto the $\xi$(magnetic field) $-$
$x$(temperature) plane for anisotropy parameter $\alpha = -1.5$
and $\xi
> 0$. The phase boundary curve is
the locus of discontinuous transitions between the longitudinal
upper phase and the transverse lower phase, and the solid circle
is a critical point associated with a continuous transition. The
reduced parameters $\xi=\tilde\mu h_z/J_3$,
$\alpha={\mathcal{D}}/J_3$, $x=T/T_c$ where
$k_BT_c/J_3=Q^{-1}_{3c}=0.25189\cdots$.}
\end{figure}

\begin{figure}
\begin{center}
\leavevmode
\includegraphics[width=0.5\textheight,angle=-90]{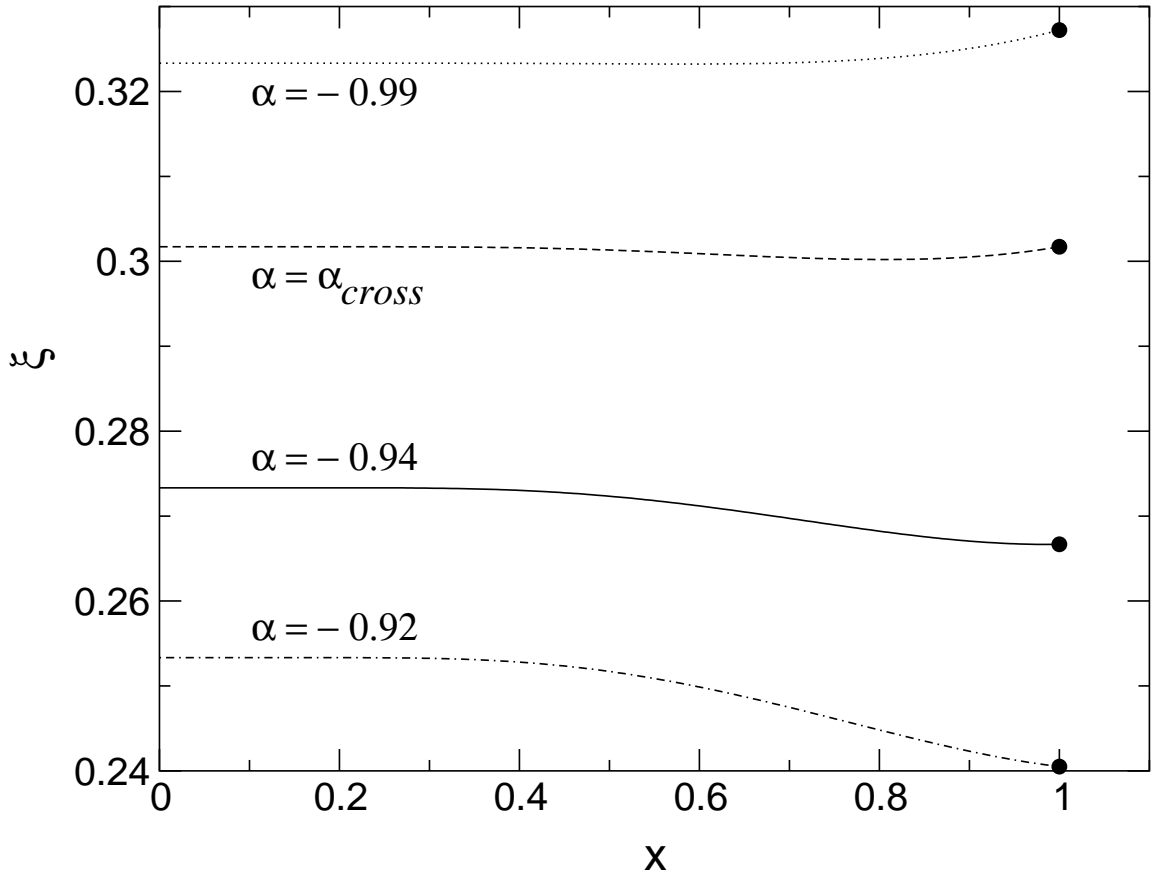}
\end{center}
\caption{The $\xi$(magnetic field) $-$ $x$(temperature) phase
boundary curves for values of the anisotropy parameter $\alpha$
where the curves change from being monotonically increasing to
monotonically decreasing, via a non-monotonic region. At
$\alpha_{cross}=-.96837\cdots$, the $x=0$ isotherm crosses the
$x=1$ isotherm in the $\xi - \alpha$ plane. The reduced parameters
$\xi=\tilde\mu h_z/J_3$, $\alpha={\mathcal{D}}/J_3$, $x=T/T_c$
where $k_BT_c/J_3=Q^{-1}_{3c}=0.25189\cdots$.}
\end{figure}

\begin{figure}
\begin{center}
\leavevmode
\includegraphics[width=0.5\textheight,angle=-90]{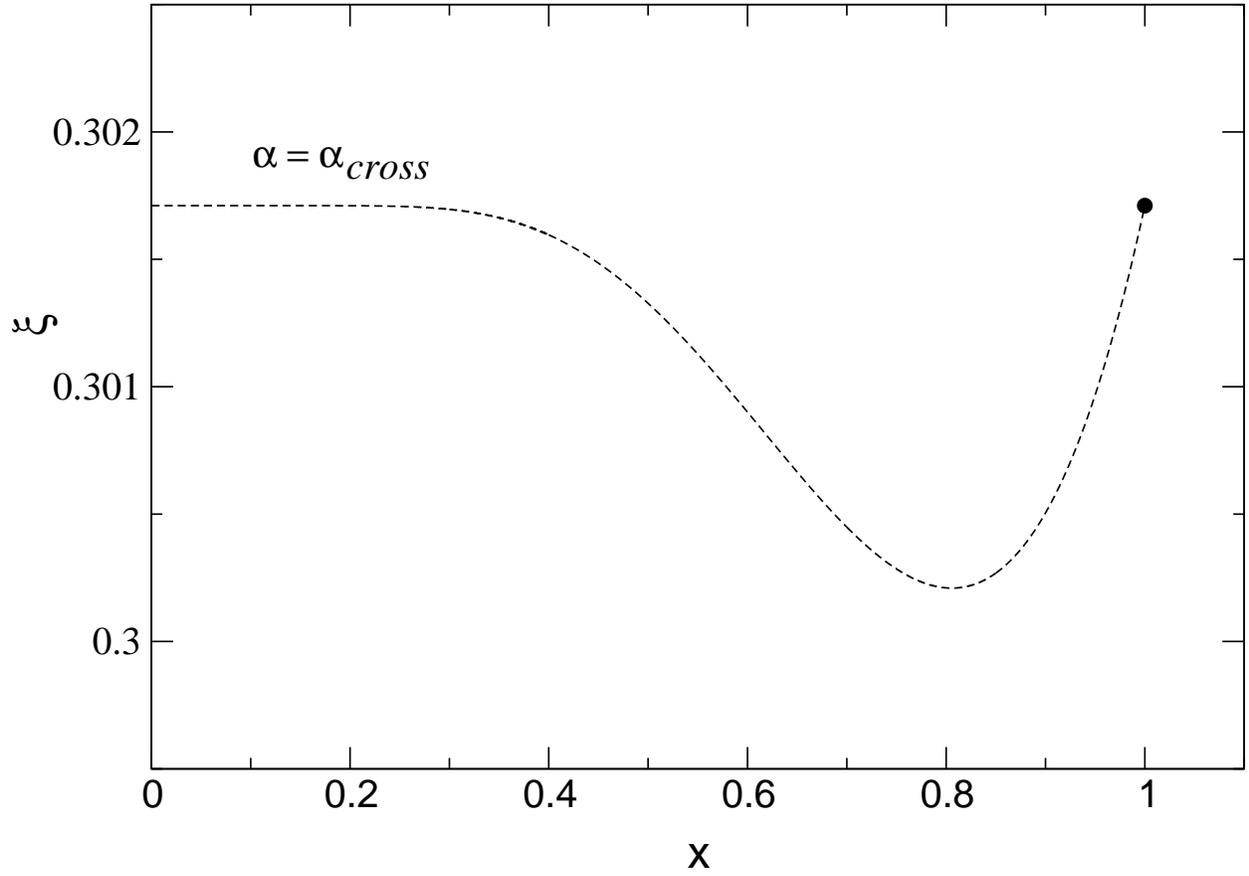}
\end{center}
\caption{The curve $\alpha =\alpha_{cross}= -0.96837\cdots$ of
Figure 6 magnified to show details of the nonmonotonicity. The
reduced parameters $\xi=\tilde\mu h_z/J_3$,
$\alpha={\mathcal{D}}/J_3$, $x=T/T_c$ where
$k_BT_c/J_3=Q^{-1}_{3c}=0.25189\cdots$.}
\end{figure}

\begin{figure}
\begin{center}
\leavevmode
\includegraphics[width=0.5\textheight,angle=-90]{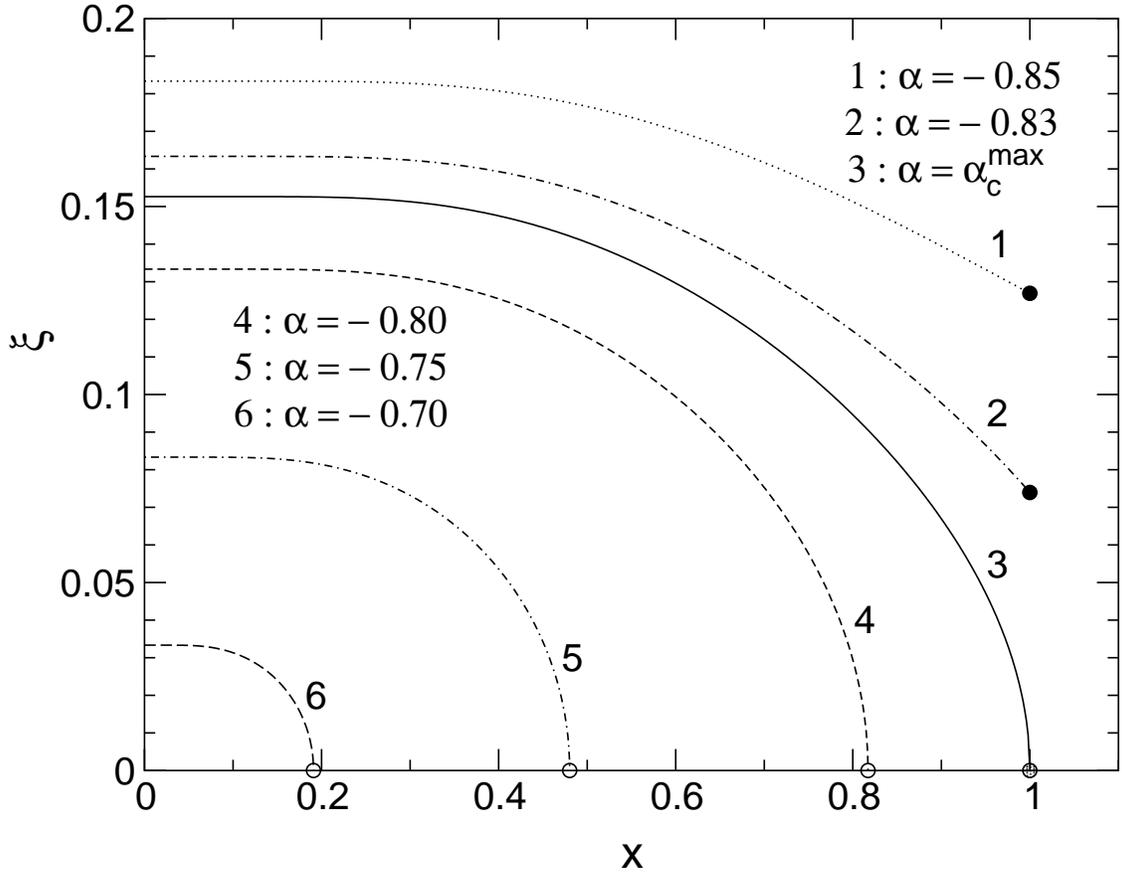}
\end{center}
\caption{The $\xi$(magnetic field) $-$ $x$(temperature) phase
boundary curves for several values of the anisotropy parameter
$\alpha$ around $\alpha^{max}_c=-0.81929\cdots$. The figure
extends symmetrically for $\xi < 0$. Curves $1$ and $2$ end at
solid circles representing critical points, while  curves $4$, $5$
and $6$ end at open circles representing confluent singularities.
For curve $3$, the critical point meets its reflective image
critical point at the confluent-critical point (shaded circle).
The reduced parameters $\xi=\tilde\mu h_z/J_3$,
$\alpha={\mathcal{D}}/J_3$, $x=T/T_c$ where
$k_BT_c/J_3=Q^{-1}_{3c}=0.25189\cdots$.}
\end{figure}

\begin{figure}
\begin{center}
\leavevmode
\includegraphics[width=0.5\textheight,angle=-90]{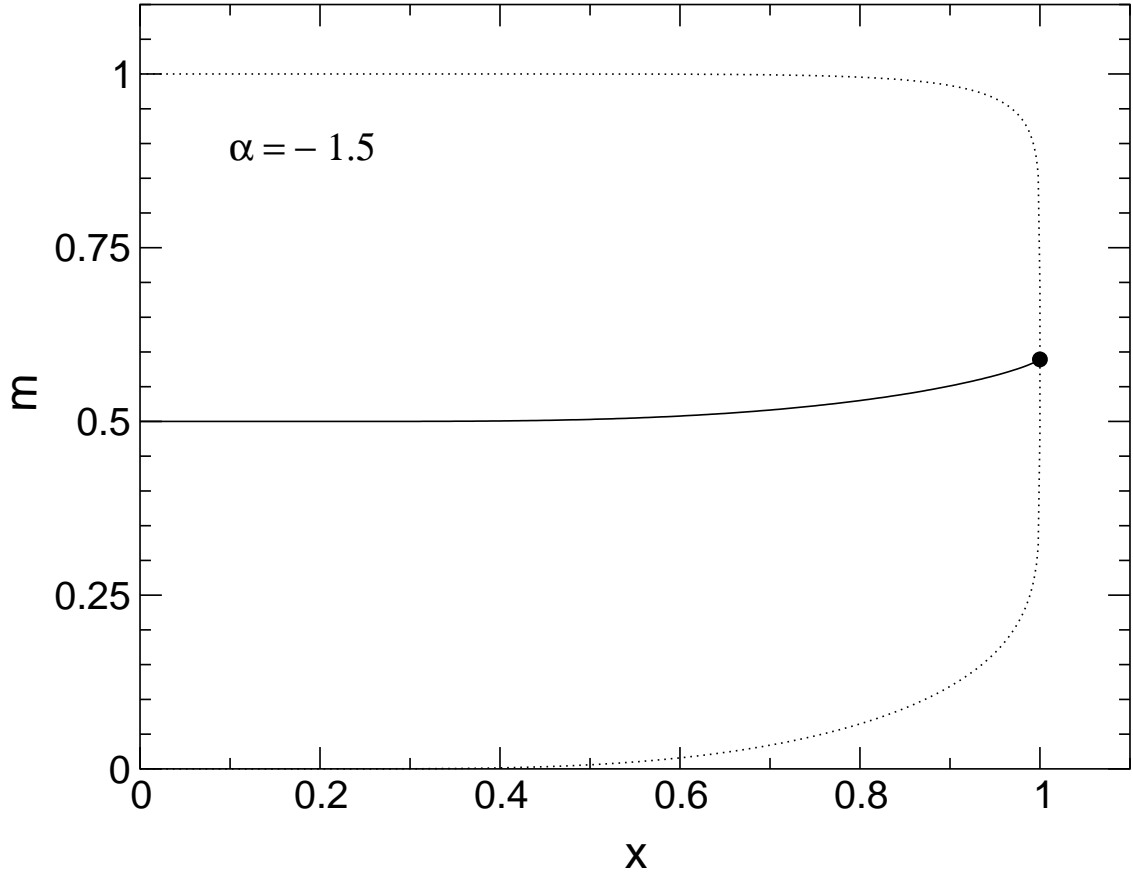}
\end{center}
\caption{The longitudinal-transverse coexistence curve (dotted
line) and the corresponding curvilinear diameter (solid line) for
$\alpha = -1.5$ in the $m$(magnetization) $-$ $x$(temperature)
plane. The coexistence curve is singular at its critical point
(solid circle). The reduced parameters $m=<S_i>$,
$\alpha={\mathcal{D}}/J_3$, $x=T/T_c$ where
$k_BT_c/J_3=Q^{-1}_{3c}=0.25189\cdots$.}
\end{figure}

\begin{figure}
\begin{center}
\leavevmode
\includegraphics[width=0.5\textheight,angle=-90]{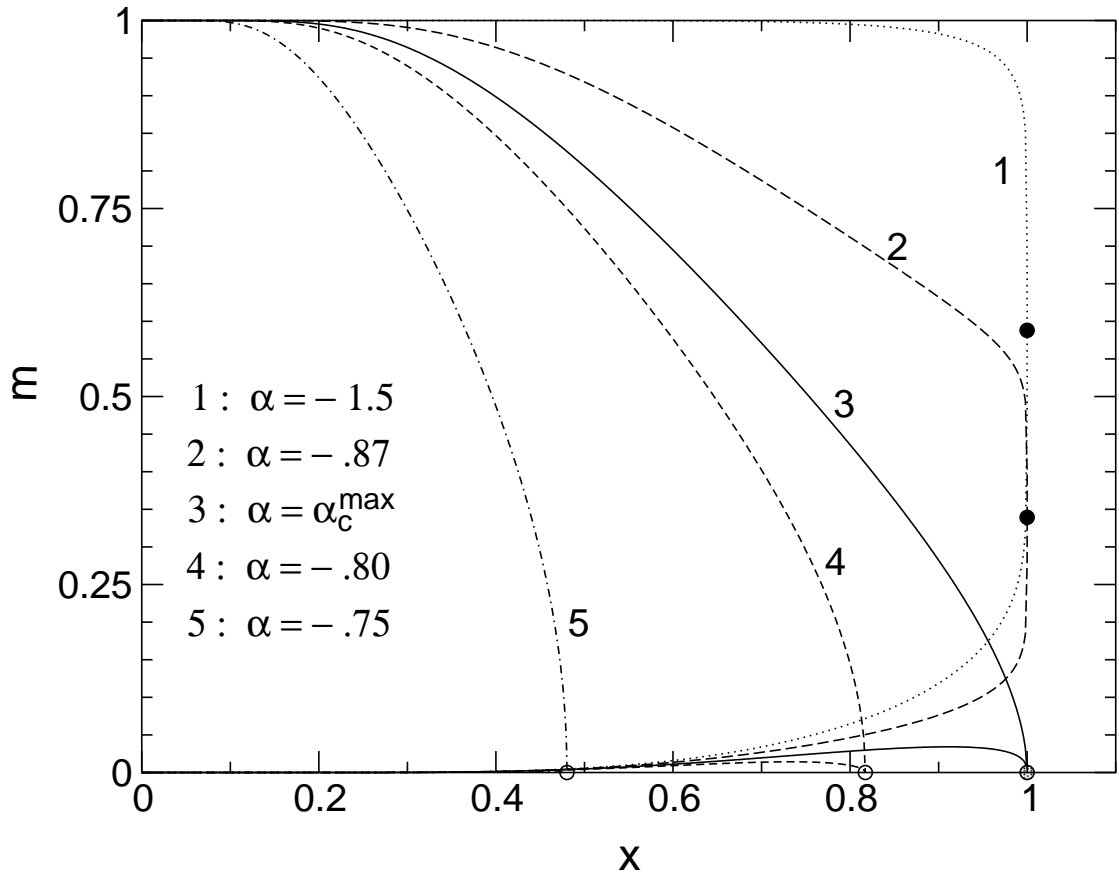}
\end{center}
\caption{Longitudinal-transverse coexistence curves in the
$m$(magnetization) $-$ $x$(temperature) plane for several values
of $\alpha$ around $\alpha^{max}_c=-0.81929\cdots$. Solid circles
are critical points, open circles are confluence points, and the
shaded circle is a confluent-critical point. The figure extends
symmetrically for $m < 0$, so that curves $3$, $4$ and $5$ entrap
homogeneous transverse phases within tadpole-shaped regions. No
such trapped transverse phases exist for curves $1$ or $2$. The
reduced parameters $m=<S_i>$, $\alpha={\mathcal{D}}/J_3$,
$x=T/T_c$ where $k_BT_c/J_3=Q^{-1}_{3c}=0.25189\cdots$.}
\end{figure}

\begin{figure}
\begin{center}
\leavevmode
\includegraphics[width=0.5\textheight,angle=-90]{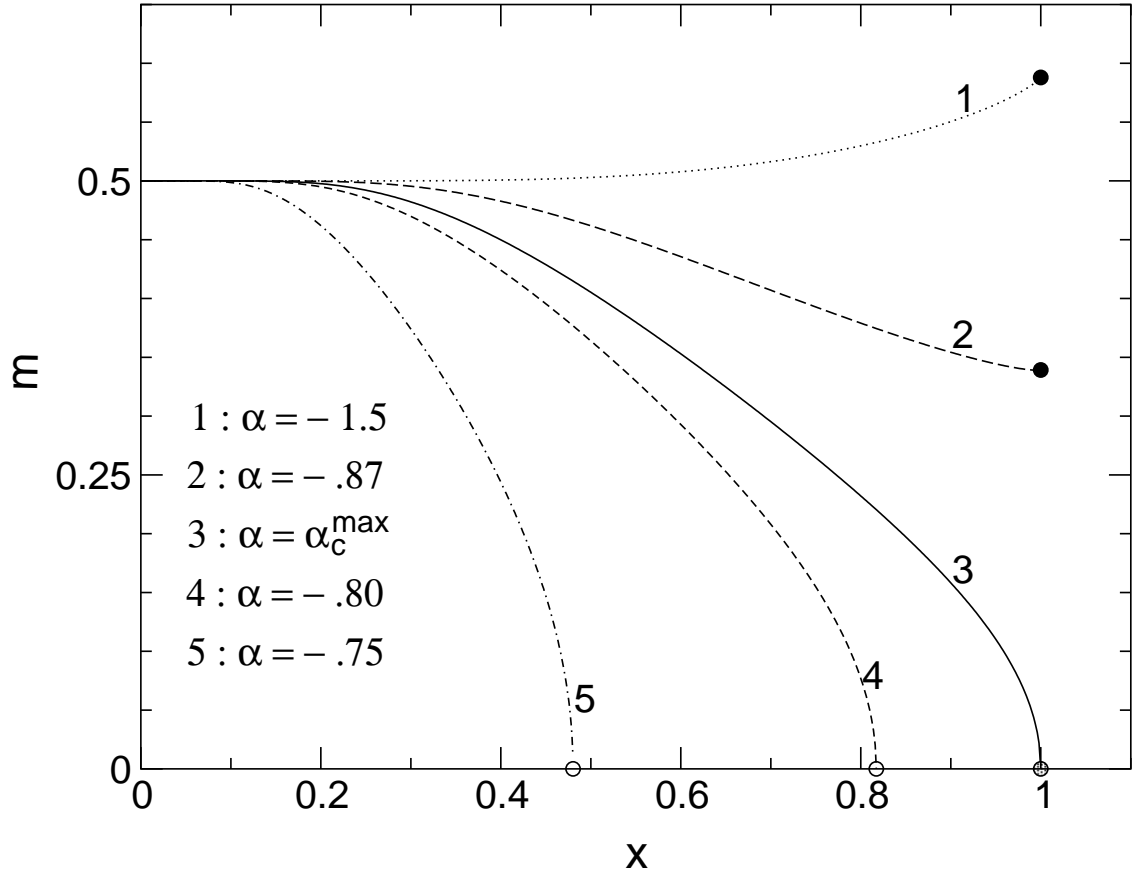}
\end{center}
\caption{Curvilinear diameters corresponding to the coexistence
curves of Figure 10, showing the diameters changing  from
monotonically increasing to monotonically decreasing behavior and
ending either at a critical point (solid circle), or eventually at
a  confluence (open circle) point or a confluent-critical (shaded
circle) point. The reduced parameters $m=<S_i>$,
$\alpha={\mathcal{D}}/J_3$, $x=T/T_c$ where
$k_BT_c/J_3=Q^{-1}_{3c}=0.25189\cdots$. The value
$\alpha^{max}_c=-0.81929\cdots$.}
\end{figure}


\begin{references}
\bibitem{BEG}M. Blume, V.J. Emery and R. B. Griffiths, Phys. Rev.
A {\bf 4}, 1071 (1971).
\bibitem{BC}M. Blume, Phys. Rev. {\bf 141}, 517 (1966); H. W.
Capel, Physica {\bf 32}, 966 (1966); {\bf 33}, 295 (1967); {\bf
37}, 423 (1967).
\bibitem{Keskin}See, e.g., M. Keskin and A. Solak, J. Chem. Phys. {\bf
112}, 6396 (2000), and references cited therein; X. N. Wu and F.
Y. Wu, J. Stat. Phys. {\bf 50}, 41 (1988).
\bibitem{Griffiths}R. B. Griffiths, Physica {\bf 33}, 689 (1967).
\bibitem{Wu}F. Y. Wu, Chinese J. Phys. {\bf 16}, 153 (1978).
\bibitem{Sullivan}N. S. Sullivan, Bull. Mag. Reson. {\bf 11}, 86 (1987) and
references cited therein; M. Roger, J. J. Hetherington and J. M.
Delrieu, Rev. Mod. Phys. {\bf 55}, 1 (1983).
\bibitem{BS}J.H. Barry and N.S. Sullivan, Int. J. Mod. Phys B {\bf 7},
2831 (1993).
\bibitem{Griffiths2}See, e.g., R. B. Griffiths, in {\it Phase
Transitions and Critical Phenomena}, vol. {\bf 1}, ed. C. Domb and
M. S. Green (Academic, New York, 1972); I. Syozi, {\it ibid}; R.
J. Baxter, {\it Exactly Solved Models in Statistical Mechanics},
(Academic, New York, 1982).
\bibitem{Houtappel}R. J. Baxter and I. G. Enting, J. Phys. A {\bf 11}, 2463 (1978);
R. M. F. Houtappel, Physica {\bf 16}, 425 (1950); S. Naya, Prog.
Theor. Phys. {\bf 11}, 53 (1954); R. J. Baxter, ref. 8; J. H.
Barry, T. Tanaka, M. Khatun and C. H. M\'{u}nera, Phys. Rev. B
{\bf 44}, 2595 (1991).

\end{references}
\end{document}